\documentclass{nature}
\usepackage{amsmath}    
\usepackage{graphicx}   
\usepackage{color}
\usepackage[10pt]{moresize}
\usepackage{graphics,graphicx}
\usepackage{amsfonts}
\usepackage{amssymb}
\usepackage{amscd}
\usepackage{amsmath}
\usepackage{enumerate}
\usepackage{epsfig}
\usepackage{subfigure}

\begin{document}

\title{Strong single-photon to two-photon bundles emission in spin-1 Jaynes-Cummings model}

\author{Jing Tang$^{1}$ and Yuangang Deng$^{1, \ast}$ }
\maketitle
\begin{affiliations}
\item
Guangdong Provincial Key Laboratory of Quantum Metrology and Sensing $\&$ School of Physics and Astronomy, Sun Yat-Sen University (Zhuhai Campus), Zhuhai 519082, China
 $^\ast$ To whom correspondence should be addressed.
E-mail: dengyg3@mail.sysu.edu.cn
\end{affiliations}

\baselineskip24pt

\maketitle

\begin{abstract}
The realization of high-quality special nonclassical states beyond strong single atom-cavity coupling regime is a fundamental element in quantum information science. Here, we study the nonclassical photon emission in a single spin-1 atom coupled to an optical cavity with constructing a spin-1 Jaynes-Cummings model. By tuning quadratic Zeeman shift, the energy-spectrum anharmonicity can be significantly enhanced with respect to the dressed-state splitting of well-resolved n-photon resonance largely increased. The photon emission exhibit high-quality single photon and two-photon bundles properties with large photon numbers in the cavity and atom driven cases, respectively. More interestingly, nonclassical optical switching from strong single-photon blockade to two-photon bundles and super-Poissonian photon emission is achieved and highly controllable by light-cavity detuning in the presence of both atom and cavity driven fields. Our proposal not only open up a new avenue for generating high-quality n-photon sources but also provide versatile applications in quantum networks and quantum metrology.
\end{abstract}

\section*{INTRODUCTION}
The interfacing of ultracold quantum gases in quantized optical cavities could offer a preeminent platform for quantum information science and studying fundamental physics~\cite{RevModPhys.85.553,RevModPhys.91.025005,mivehvar2021cavity}. In particular, the manipulation of nonclassical states of quantum light at single (few) photon levels is an important building block in quantum communication~\cite{Duan01}, quantum computation~\cite{Bennett00}, quantum cryptography~\cite{RevModPhys.74.145}, and other emerging quantum technologies~\cite{Giovannetti2011, Georgescu14, Brien09}. Recent experiments involving a single atom cavity quantum electrodynamics (QEDs) have implemented a high-quality deterministic single-photon sources mediated by enhancing the light-matter interaction~\cite{RevModPhys.73.565, RevModPhys.82.1041,RevModPhys.87.1379}.

As to single photons, there have two paradigmatic physical mechanisms by utilizing the conventional photon blockade (PB)~\cite{PhysRevLett.79.1467} and unconventional PB~\cite{Liew10,Tang15}, where the photon emission hosts a sub-Poissonian distribution and antibunching quantum statistics since excitation of one photon will block or eliminate a second photon. Up to now, these two mechanisms as the key techniques for generating controllable single-photon state have been extensively studied. The conventional PB utilizing the large enough strong energy-spectrum anharmonicity of optical nonlinearity, has been theoretical
proposed and experimental advanced in a sequence of setups including atom-cavity QEDs~\cite{Birnbaum2005,Fink08, Tang21}, artificial atoms-cavity system~\cite{Hennessy07, Reinhard2012, Kai15}, optomechanical resonators~\cite{Rabl11,Nunnenkamp11}, waveguide or circuit QEDs~\cite{Liu14,Hoffman11}, and Kerr-type nonlinear systems~\cite{PhysRevLett.121.153601, PhysRevLett.123.013602}. On the other hand, the unconventional PB relies on the construction of quantum destructive interference between different quantum transition pathways~\cite{Bamba2011,Majumdar2012,PhysRevLett.121.043601,li2019nonreciprocal, Tang19}. Therefore the two-photon excitation can be completely eliminated by adding an auxiliary driving field~\cite{Tang19}, multimode cavity field~\cite{PhysRevLett.127.240402,PhysRevA.105.043711} or qubit~\cite{PhysRevA.100.063817, Zhu21}, which can mitigate the strong coupling condition of conventional PB.

In contrast to single-PB,  $n$-photon bundles state which possesses special statistic properties with bunching for single photons and antibunching for separated bundles of photons could offer new resources for exploring intriguing fundamental physics and applications in quantum technologies~\cite{afek2010high, RevModPhys.88.045008}. Remarkably, the $n$-quanta bundles state by utilizing the Mollow physics~\cite{munoz2014emitters,munoz2018filtering}, deterministic parametric down-conversion~\cite{PhysRevLett.117.203602}, Stokes or parity-symmetry-protected multiphoton process~\cite{PhysRevLett.124.053601,PhysRevLett.127.073602}, and $n$-phonon resonance process~\cite{Deng:21} have been proposed. In these significant advances, the high-purity $n$-quanta sources operating at high rates in the far dispersive or strong coupling regime remains a challenge in experiments. Meanwhile, we should note that the conventional PB mechanism is hard to be extended to study $n$-quanta bundles state with building the complicated quantum interference conditions.

Originally these phenomena including single-PB and $n$-photon bundles emission are focused on a simplified Jaynes-Cummings model (JCM) that describes a single two-level atom (or equivalent spin-$1/2$ degrees of freedom) interacting with a quantized electromagnetic single-mode optical (acoustic) cavity~\cite{RevModPhys.93.025005,haroche2006exploring}. An interesting question is whether the high-quality $n$-photon emissions in high-spin degrees of freedom systems, which is beyond the standard two-level JCM with requiring strong atom-cavity coupling, can be generated. An affirmative answer will not only enrich our knowledge of the fundamental nature of seminal quantum model in high-spin systems but also provide a broad physics community for applications in quantum physics~\cite{PhysRevLett.127.033602, zhiqiang2017nonequilibrium, PhysRevLett.122.103601, PhysRevLett.127.253601, PhysRevLett.128.153601}.

In this work, we propose to realize high-quality nonclassical photon emission from single-PB to two-photon bundles state for a single spin-1 atom trapped in a high-finesse optical cavity. Compared to the standard two-level JCM, the dressed-state splitting of $n$-photon resonance is significantly increased by tuning the quadratic Zeeman shift in spin-1 JCM. Remarkably, the corresponding strong energy-spectrum anharmonicity emerges beyond the conventional strong coupling regime  as a result of versatile spin degrees of freedom. We show that, with driving the cavity (atom) field, high-quality single-photon (two-photon bundles) emission is observed, corresponding to the large steady-state photon numbers concurrently. In particular, the generated second-order correlation functions for single-photon and two-photon bundle states can be low to $g_1^{(2)}(0)=7.5\times 10^{-5}$ and $g_2^{(2)}(0)=4.9\times 10^{-5}$ at a moderate single atom-cavity coupling, which indicate the strong antibunching for isolated photons and separated two-photon bundles, respectively. Furthermore, the photon emission exhibits strong single-PB to two-photon bundles and super-Poissonian photon state by tuning the light-cavity detuning with interplay of the cavity driven and atom pump fields, which will facilitate the versatile technological applications such as nonclassical optical switchings and high-quality multiphoton sources~\cite{PhysRevLett.127.133603, xu18, Ren2019, Rempe2022, Pan2022}. Our study will motivate the relevant studies in exploring $n$-photon bundles emission in high-spin JCM and significantly enhance our understanding on fundamental nature of the model.

\section*{RESULTS}
\subsection{Model and energy spectrum.}
\begin{figure}
\centering \resizebox{12cm}{!}{\includegraphics{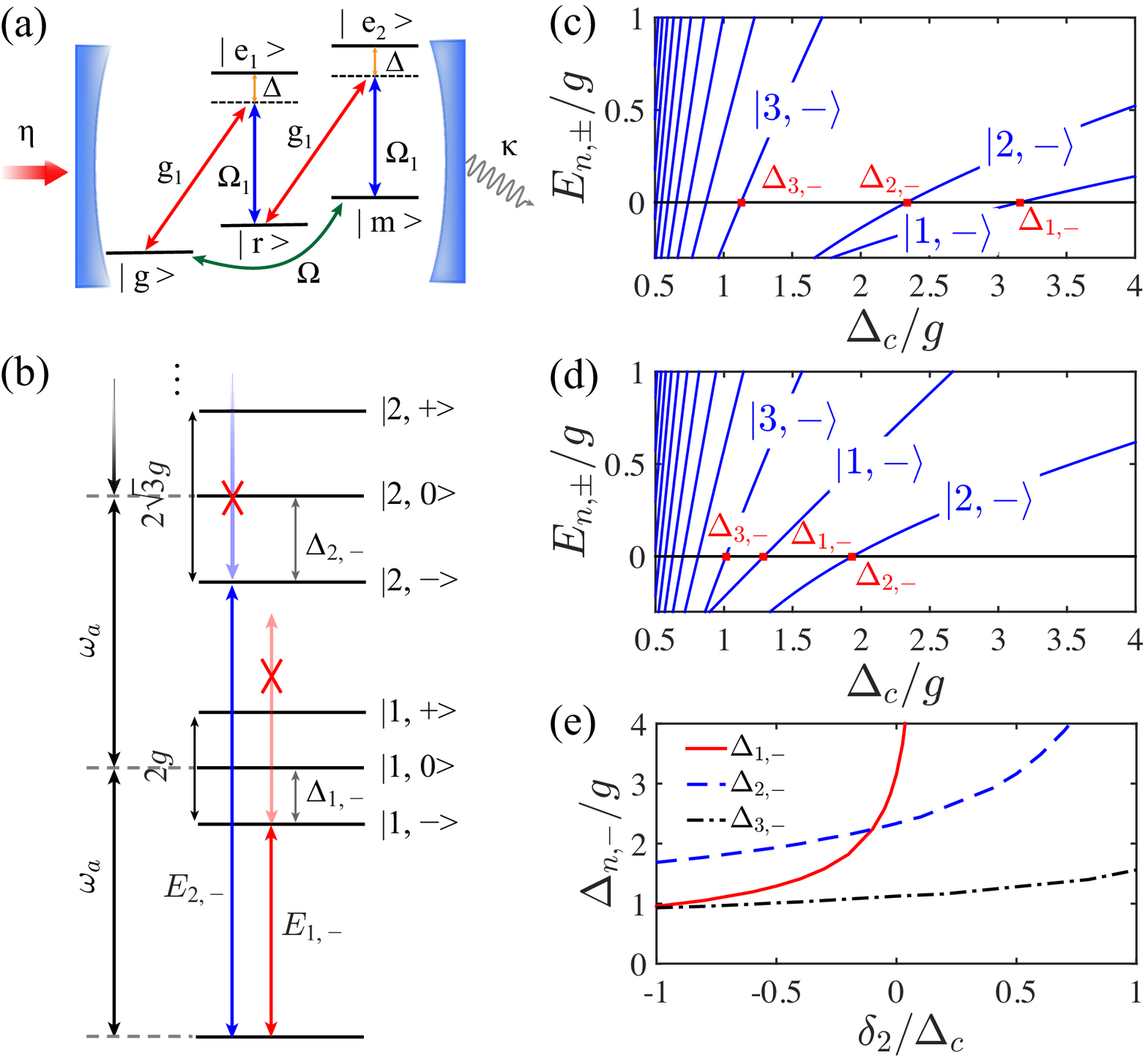}}
\caption{(color online). {\bf Schematic for creating the high-quality $n$-photon sources.} (a)  A single atom trapped in an optical cavity with constructing a spin-1 JCM. (b) The typical anharmonicity ladder of the energy spectrum for $\delta_1=\Delta_c$ and $\delta_2=0$. (c)-(d) The energy spectrum of Hamiltonian (\ref{Hamiltonian2}) for $\delta_2/\Delta_c=0$ and $\delta_2/\Delta_c=-0.5$, respectively. The red square marks the position of  $n$-photon resonance $\omega_n$. (e) The $n$-photon resonance $\omega_n$ as a function of $\delta_2$. In (c)-(e), the other parameter is $\delta_1/\Delta_c=0.1$.}  \label{scheme}
\end{figure}

We consider a single atom subjected to a high-finesse optical cavity, as depicted in Fig.~\ref{scheme}(a). The relevant energy levels include two electronic excited states $|e_1\rangle$ and $|e_2\rangle$, and three internal atomic states from the pseudospin-$1$ ground electronic manifold $|g\rangle$, $|r\rangle$, and $|m\rangle$. In our configuration, the atomic dipole transitions $|g\rangle \leftrightarrow |e_1\rangle$ and $|r\rangle \leftrightarrow |e_2\rangle$ are far-off resonant with the single-mode cavity field, corresponding to the single atom-cavity coupling strength $g_1$ and large atom-cavity detuning $\Delta$. To generate Raman couplings, a classical pump field with Rabi frequency $\Omega_1$ is applied to drive the atomic transitions $|r\rangle \leftrightarrow |e_1\rangle$ and $|m\rangle \leftrightarrow |e_2\rangle$. We should emphasize that the quadratic Zeeman effect within the pseudospin-$1$ ground manifold should not be too large to realize  resonance Raman coupling processes. This approximation can be well satisfied by subjecting to a weak bias magnetic field ${\bf B}$ along the cavity axis (quantization axis). In addition, the optical cavity with decay rate $\kappa$ is driven by a weak laser field with single-photon parametric drive  amplitude $\eta$, and cavity-light detuning $\Delta_c'$. Furthermore, the two ground states $|g\rangle$ and $|m\rangle$ assume coupled by an external two-photon Raman coupling with a weak Rabi frequency $\Omega$. As we shall see below, the atom pump field plays an essential role in realizing two-photon bundles state.

The far-off resonance excited states $|e_1\rangle$ and $|e_2\rangle$ can be adiabatically eliminated in the far dispersive regime, i.e., $|g_1/\Delta|\ll1$ and $|\Omega_1/\Delta|\ll1$. Under the rotating-wave approximation, the interaction Hamiltonian of single atom-cavity system is given by
\begin{align}\label{Hamiltonian1}
\hat{\cal H}_{I}/\hbar&= \Delta_c \hat{a}^\dag\hat{a}+ \Delta_r \hat{\sigma}_{rr}+ \Delta_m \hat{\sigma}_{m}+g(\hat{a}^\dag   \hat{\sigma}_{gr}+\hat{a} \hat{\sigma}_{rg})    \nonumber \\
&+g(\hat{a}^\dag  \hat{\sigma}_{rm}+\hat{a} \hat{\sigma}_{mr}),
\end{align}%
where $\hat{a}^\dag$ and $\hat{a}$ are the creation (annihilation) operator of cavity mode and $\hat{\sigma}_{ij}=|i\rangle  \langle j|$ is the atomic spin projection operators with $i, j=[{g, r, m}]$ labeling the atomic internal states. Here $g=-g_1\Omega_1/\Delta$ is the effective single atom-cavity coupling reduced by the two-photon Raman process, $\Delta_c=\Delta_c' - g_1^2/\Delta$ is the effective cavity-light detuning, and $\Delta_r$ ($\Delta_m$) is the tunable one- (two-) photon detuning of the atom field.

To gain more insight into the system, the Hamiltonian (\ref{Hamiltonian1}) can be rewritten in the spin-$1$ representation
\begin{align}\label{Hamiltonian2}
\hat{\cal H}_{I}/\hbar&= \Delta_c \hat{a}^\dag\hat{a} + \delta_1 \hat{S}_z+ \delta_2 \hat{S}_z^2+\frac{g}{\sqrt{2}}(\hat{a}^\dag  \hat{S}_{-}+ \hat{a} \hat{S}_{+}),
\end{align}%
where $\hat{S}_{x,y,z}$ are spin-1 matrices and  $\delta_1=\Delta_m/2$ ($ \delta_2=\Delta_m/2-\Delta_r$) is the effective linear (quadratic) Zeeman shift. Explicitly, the angular momentum operators are defined by: $\hat{S}_z= \hat{\sigma}_{mm}-\hat{\sigma}_{gg}$ and $\hat{S}_-=\hat{S}_x-i\hat{S}_y=\sqrt{2} (\hat{\sigma}_{gr}+\hat{\sigma}_{rm})$ with $\hat{S}_+=\hat{S}_-^\dag$. Remarkably, the Hamiltonian (\ref{Hamiltonian2}) essentially characterize a spin-$1$ JCM, which is in contrast to the seminal JCM of describing a single two-level atom (qubit) coupling to a quantized single-mode optical (acoustic) cavity~\cite{RevModPhys.93.025005,haroche2006exploring}. In particular, the spin-$1$ JCM of Eq.~(\ref{Hamiltonian2}) possesses a continuous $U(1)$ symmetry characterized by the operator ${\cal R}_{\theta} =\exp[i\theta(\hat{a}^\dag\hat{a}+\hat{S}_z)]$, which yields ${\cal R}_{\theta}^\dag(\hat{a},\hat{S}_-
,\hat{S}_+){\cal R}_{\theta} =(\hat{a} e^{-i\theta},\hat{S}_-e^{-i\theta},\hat{S}_+e^{i\theta})
$. Therefore, a gapless zero-energy Goldstone mode of low-energy excitation can be expected for single-atom superradiance~\cite{science.aar2179}, corresponding to the continuous $U(1)$ symmetry spontaneous breaking~\cite{PhysRevLett.112.173601}.

To investigate the underlying physical mechanism of multiphoton emission in the spin-1 JCM, it is instructive to calculate the energy spectrum of the system by diagonalizing Hamiltonian (\ref{Hamiltonian2}). In the absence of atom and cavity dissipations and driven fields, the total excitation number operator $\hat{N}= \hat{a}^\dag\hat{a} + \hat{S}_z +1$ in our system is conserved, which satisfies the commutation relation $[\hat{\cal H},{\hat{N}}] = 0$. Fixing the excitation numbers of the cavity as $n$, the accessible subspace of Hilbert space is restricted to Fock states with respect to the relevant states $|n, g\rangle$, $|n-1,r\rangle$, and $|n-2,m\rangle$. As a consequence, the corresponding matrix ${\mathcal {M}}$ can be readily obtained by solving the Schr$\ddot{o}$dinger equation, ${\hat{\cal H}}\Psi={\cal M}\Psi$, with $\Psi=[|n, g\rangle, n-1,r\rangle, |n-2,m\rangle]^T$. Explicitly, the matrix ${\mathcal {M}}$ takes the form as
\begin{align}\label{matrix}
{\mathcal {M}}=\left(\begin{array}{ccc}
n\Delta_c& \sqrt{n}g& 0\\
 \sqrt{n}g&(n-1)\Delta_c+\delta_1-\delta_2 & \sqrt{n-1}g\\
0 & \sqrt{n-1}g & (n-2)\Delta_c+2\delta_1\\
\end{array}\right),
\end{align}%
We calculate the energy spectrum by diagonalizing the matrix of Eq.~(\ref{matrix}). The corresponding $n$th ($n>1$) dressed state split into three branches in pseudospin-1 ground manifold. Fixing $\delta_1=\Delta_c$, the typical energy eigenvalues of system read
\begin{align}
E_{n, \pm} &=n\Delta_c-\delta_2/2 \pm\sqrt{(2n-1)g^2+\delta_2^2/4}, \nonumber \\
E_{n, 0} & = n\Delta_c, \nonumber
\end{align}%
which correspond to the energy splitting for the dressed states of the higher- ($|n, +\rangle$), middle- ($|n, 0\rangle$), and lower- ($|n, -\rangle$) branches, as displayed in Fig.~\ref{scheme}(b). For $\Delta_c=0$, the middle-branch splitting $E_{n, 0}=0$ reduces to an atomic dark-state
\begin{align}\label{DS}
|n, 0\rangle=\sqrt{\frac{n}{2n-1}}|n, g\rangle  -\sqrt{\frac{n-1}{2n-1}}|n-2,m\rangle,
\end{align}%
which exists only for the multiphoton emissions ($n\geq 2$). It is clear that the $n$-photon resonance with respective to $n$th dressed states satisfy
\begin{align}
\Delta_{n,\pm} &= \left[\delta_2/2 \mp \sqrt{(2n-1)g^2+\delta_2^2/4}\right]/n,
\end{align}%
where the higher- (lower-) branch splitting increases with increasing the negative (positive) value of quadratic Zeeman shift $\delta_2$. Thus the anharmonic ladder of energy spectrum for the $n$-photon dressed states can be further enhanced by tuning $\delta_2$.

For arbitrary constants of the detuning ratios $\delta_1/\Delta_c$ and $\delta_2/\Delta_c$, the exact energy spectrum of Eq.~(\ref{matrix}) can be calculated analytically, but the expression is too bulky to be presented here. In this case, we check that a symmetric structure in the $n$th dressed-state splitting between red and blue light-cavity detunings is observed, corresponding to $n$-photon resonance satisfying $\Delta_{n,-}= -\Delta_{n,+}$. In addition, the $n$th dressed state for the middle branch maintains the dark-state in Eq.~(\ref{matrix}) with $E_{n, 0}=0$ at $\Delta_c=0$.

To further explore the energy spectrum, we plot the typical energy spectrum for different values of $\delta_2$ with fixing $\delta_1/\Delta_c=0.1$, as shown in Figs.~\ref{scheme}(c) and \ref{scheme}(d). As expected, the $n$-photon resonance occurs at the frequency $\Delta_{n,-}$ (red square) by tuning the light-cavity detuning at the red sideband ($\Delta_c>0$). Surprisingly, the two-photon resonance frequency is larger than that of single-photon resonance, i.e., $\Delta_{2,-}>\Delta_{1,-}$,  for $\delta_2/\Delta_c=-0.5$, which is in contrast to the conventional two-level JCM that always exhibit $\Delta_{1,-}>\Delta_{2,-}$ with $\Delta_{n,-}=g/\sqrt{n}$ in the atom-cavity resonance regime~\cite{Deng:21}. Ignoring the cavity driven and atom pump fields ($\eta=0$ and $\Omega=0$), we calculate the $n$-photon resonance $\Delta_{n,-}$ with respective to $\delta_2$ at $\delta_1/\Delta_c$=0.1  in Fig.~\ref{scheme}(e). Interestingly, the $\delta_2$-dependence of $n$-photon resonance $\Delta_{n,-}$ is monotonically increased. In particular, the single-photon resonance $\Delta_{1,-}$ significantly enhanced compared with two-photon resonance $\Delta_{2,-}$ with $\delta_2$ increasing. This leads to the energy level crossing at the single- and two-photon resonances when the quadratic Zeeman shift $\delta_2/\Delta_c$=-0.1. It should be noted that the quadratic Zeeman shift plays an important role in realizing nonclassical $n$-photon sources due to the obviously enhanced energy spectrum anharmonicity.

\subsection{Strong photon blockade for single photons.}
\begin{figure}
\centering \resizebox{12cm}{!}{\includegraphics{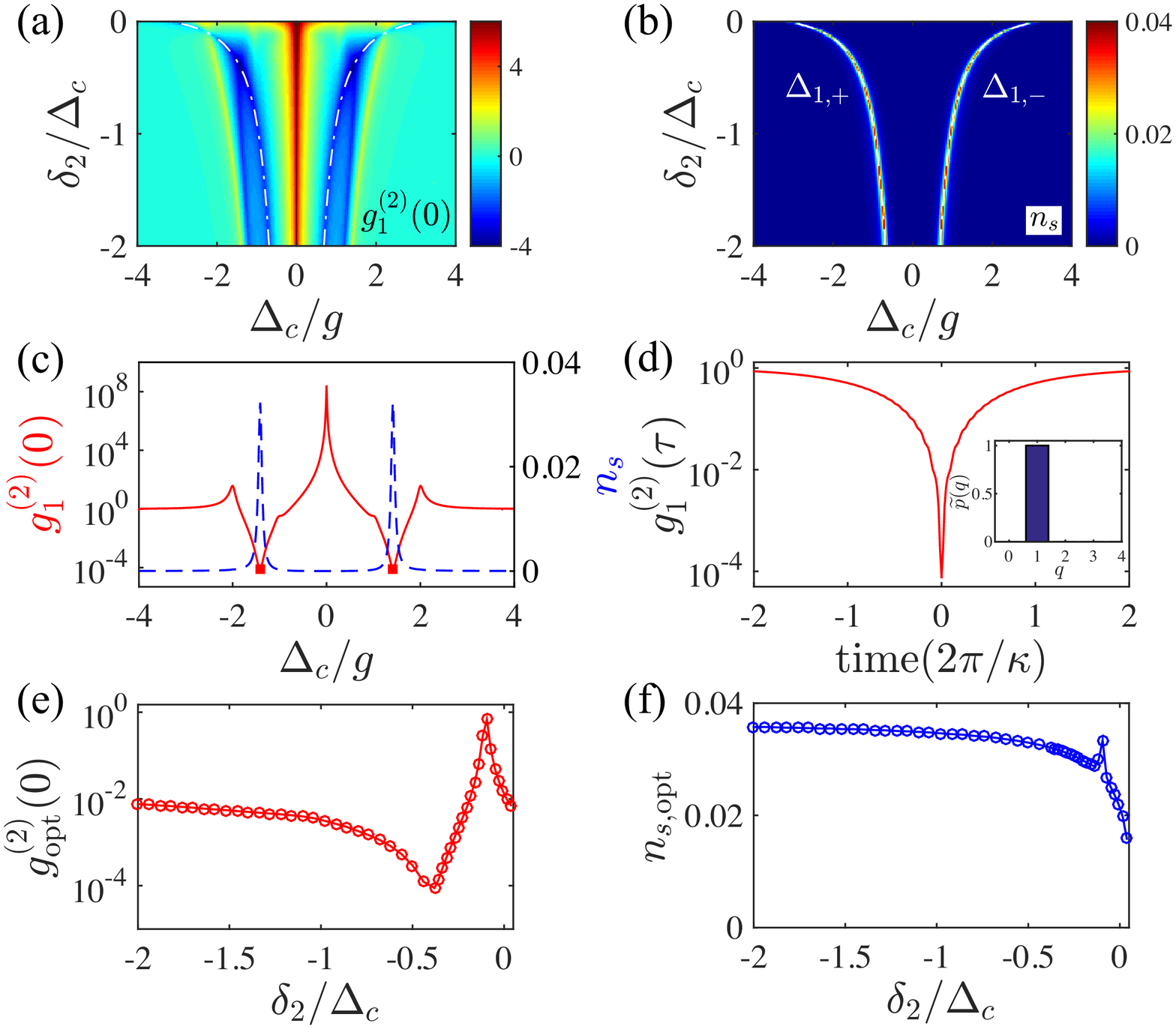}}
\caption{{\bf The photon quantum statistical properties for spin-1 JCM with the cavity driven.} Distribution of $g_1^{(2)}(0)$ (a) and $n_s$ (b) on the $\Delta_c$-$\delta_2$ parameter plane. (c) $g^{(2)}_1{(0)}$ (solid line) and the corresponding $n_s$ (dashed line) as a function of $\Delta_c$ for $\delta_2/\Delta_c=-0.4$. (d) Time interval $\tau$ dependence of $g_1^{(2)}(\tau)$ for $\delta_2/\Delta_c=-0.4$ and $\Delta_c/g=\pm1.41$. (e) The optimal $g^{(2)}_{\rm{opt}}(0)$ and (f) the corresponding $n_s,_{\rm{opt}}$ as a function of $\delta_2$, respectively. The white dashed-dotted lines in (a) and (b) show the corresponding vacuum Rabi splitting $\Delta_{1,\pm}$ for the first dressed state of energy spectrum. The colors with blue-red gradient shading represent the values of log$[g^{(2)}_1(0)]$ in (a) and $n_s$ in (b). The inset in (d) plots the typical steady-state photon-number distribution $\tilde{p}(q)$. The other parameters are $\eta/\kappa=0.1$ and $\Omega/\kappa=0$.} \label{drivencavity}
\end{figure}

To study the PB for single-photon emission, we first consider the spin-1 system in the presence of cavity driven field with fixing $\eta/\kappa=0.1$ and $\Omega=0$. We check that the energy spectrum is slightly distorted for sufficiently weak driven amplitude of cavity ($\eta/g\ll 1$). Figures~\ref{drivencavity}(a) and ~\ref{drivencavity}(b) display the numerical results for the equal time second-order correlation function $g_1^{(2)}(0)$ and steady-state cavity photon number $n_s$ as a function of light-cavity detuning $\Delta_c$ and quadratic Zeeman shift $\delta_2$. As can be seen, the single-PB exists in a large parameter regime when $\Delta_c$ and $\delta_2$ are tuned. In particular, the strong single-PB  with $g^{(2)}_1(0)<0.01$ is reached around the vacuum Rabi splitting $\Delta_{1,\pm}$ for higher and lower branches of the system (dashed-dotted lines). More importantly, obvious steady-state photon emissions for cavity field are observed in these regions as well, which are highly consistent with the analytic result of single-photon resonance in Fig.~\ref{scheme}(e). Interestingly, both $g_1^{(2)}(0)$ and $n_s$ host the symmetric structures at the red and blue sidebands of light-cavity detuning ($\Delta_{1,-}=-\Delta_{1,+}$), in spite of the dissipations and cavity driven field are fully included.

Figure~\ref{drivencavity}(c) shows the $\Delta_c$ dependence of $g_1^{(2)}(0)$ (solid line) and $n_s$ (dashed line) for $\delta_2/\Delta_c=-0.4$. Particularly, the photon emissions reach the strong sub-Poissonian statics with $g_1^{(2)}(0)=7.5\times10^{-5}$ at the single-photon resonance $\Delta_{1,\pm}\approx \pm\sqrt{2}g$, corresponding to a relatively large intracavity photon number with $n_s>0.03$. In addition, the photon emission at the middle branch ($\Delta_c/g=0$) disappears due to the dark-state in Eq.~(\ref{DS}) which requires at least the two-photon excitations ($n\ge2$). In Fig.~\ref{drivencavity}(d), we plot the typical interval dependence of second-order correlation function $g_1^{(2)}(\tau)$ at $\delta_2/\Delta_c=-0.4$ and $\Delta_c/g=\pm1.41$. Clearly, the strong single-PB exhibiting sub-Poissonian distribution with $g_1^{(2)}(0)<10^{-4}$ and photon antibunching with $g_1^{(2)}(0)<g_1^{(2)}(\tau)$ is confirmed. To quantify the quality of the single-photon emission, we introduce the fraction of $q$-photon states to total excitation photons, $\tilde{p}(q)=qp(q)/n_s$, where $p(q)={\rm tr}(|q\rangle\langle q| \rho_s)$ measures the steady-state photon-number distribution. As shown in the inset of Fig.~\ref{drivencavity}(d), the photon emission for the single-photon nature reaches to nearly 100$\%$ at $\Delta_c=\Delta_{1,\pm}$. We point out that the decay of the antibunching  for single photons is proportional to $1/\kappa$, which characterizes the lifetime of single-photon emission.

For quantitatively characterizing the influence of $\delta_2$ on the quantum statistics, we plot the optimal $g^{(2)}_{\rm{opt}}(0)=\min [g_1^{(2)}(\Delta_c)]$ and the corresponding photon number $n_s,_{\rm{opt}}$ at the same light-cavity detuning as a function of $\delta_2$, as shown in Figs.~\ref{drivencavity}(e) and ~\ref{drivencavity}(f). With $\delta_2$ increasing, the value of $g^{(2)}_{\rm{opt}}(0)$ monotonically decreases, until it reach the dip with $g^{(2)}_{\rm{opt}}(0)=7.5\times 10^{-5}$ at $\delta_2/\Delta_c=-0.4$, and then  it rapidly increases for $\delta_2/\Delta_c < -0.1$. Moreover, although the $n_s,_{\rm{opt}}$ decreases with the increasing $\delta_2$, the corresponding $n_s,_{\rm{opt}}$ is still relatively large in the strong single-PB regime, which is conducive to generating high-quality single-photon sources. We have checked that  $g^{(2)}_1(0)=7.0\times 10^{-2}$ at $\Delta_c/g=\pm1$ in two-level JCM with the same parameters as spin-1 JCM, which proved that the anti-bunching of PB in spin-1 JCM can be improved by three orders of magnitude. The physical mechanism in spin-1 JCM for generating strong single-PB can be understood by examining the strong energy-spectrum anharmonicity of $n$-photon dressed states. Interestingly, the single-photon resonance $\Delta_{1,-}$ will dramatically increase by tuning the quadratic Zeeman shift $\delta_2$. However, the single-photon and two-photon resonances $\Delta_{1,-}$ and $\Delta_{2,-}$ will merge to a single point at $\delta_2/\Delta_c$=-0.1 [Fig.~\ref{scheme}(e)]. As a result, the single-PB is suppressed at the level crossing between single-photon and two-photon resonance due to occurring of the two-photon excitation.

\subsection{High-quality two-photon bundles emission.}
\begin{figure}
\centering \resizebox{12cm}{!}{\includegraphics{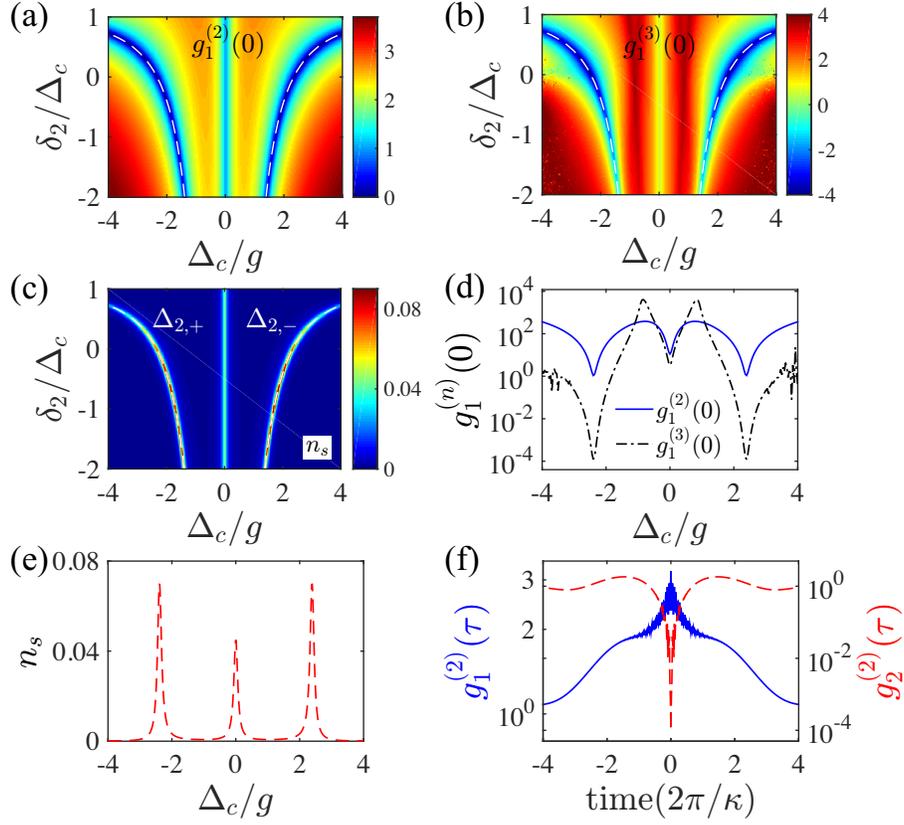}}
\caption{(color online). {\bf The photon quantum statistical properties for spin-1 JCM with the atom pump field}. Distributions of (a) $g_1^{(2)}(0)$, (b) $g_1^{(3)}(0)$, and (c) $n_s$ as functions of $\Delta_c$ and $\delta_2$. The white dashed lines show the analytical two-photon resonance $\Delta_{2,\pm}$ for the second dressed states of energy spectrum. (d) $g_1^{(2)}(0)$ (solid line) and $g_1^{(3)}(0)$ (dashed-dotted line), and (e) the corresponding $n_s$ as a function of $\Delta_c$ with $\delta_2/\Delta_c=0.05$. (f) Time interval $\tau$ dependence of  $g_1^{(2)}(\tau)$ (solid line) and $g^{(2)}_{2}(\tau)$ (dashed line) for $\Delta_c/g= 2.5$ and $\delta_2/\Delta_c=0.05$. The other parameters are fixed at $\eta/\kappa=0$ and $\Omega/\kappa=0.08$. }
\label{drivenatom}
\end{figure}

Now we focus on the two-photon bundles emission by employing a weak atom pump field with fixing $\eta=0$ and $\Omega/\kappa=0.08$. Figures~\ref{drivenatom}(a) and \ref{drivenatom}(b) illustrate the photon quantum statistics of log[$g_1^{(2)}(0)$] and log[$g_1^{(3)}(0)$]  in the $\Delta_c$-$\delta_2$ parameter plane, respectively. In the regime of two-photon resonance $\Delta_{2,\pm}$ (dashed lines), the photon statistics of second- and three-order correlation functions satisfy $g_1^{(2)}(0)>1$ and $g_1^{(3)}(0)<1$, which demonstrate that the two-PB satisfies two-photon super-Poissonian distribution and three-photon sub-Poissonian distribution simultaneously. Interestingly, $g_1^{(3)}(0)$ decreases significantly with the  increasing $\delta_2$, corresponding to the energy splitting for two-photon resonance is rapidly enhanced [Fig.~\ref{scheme}(e)]. Fig.~\ref{drivenatom}(c) displays the corresponding $n_s$ with the same parameters as Figs.~\ref{drivenatom}(a) and \ref{drivenatom}(b). Besides the red and blue sidebands at the two-photon resonance ($\Delta_{2,-}=-\Delta_{2,+}$), the $\delta_2$-independent middle branch of the dark-state also exhibits a large photon emissions at $\Delta_c=0$. The three-peak profile of $n_s$ for nonzero atom pump field is essential different from the two-peak profile with $\Omega=0$ as displayed in Fig~\ref{drivencavity}(b).

In Figs.~\ref{drivenatom}(d) and \ref{drivenatom}(e), we plot the $\Delta_c$ dependence of $g_1^{(n)}(0)$ and $n_s$ for $\delta_2/\Delta_c=0.05$, respectively. As can be seen, the strong PB for two-photon emission exhibiting super-Poissonian distribution with $g_1^{(2)}(0)=1.1$ and $g_1^{(3)}(0)=1.2\times10^{-4}$ can be achieved. Moreover, the maximal photon emission $n_s$ is also observed at  both the red and blue sidebands of the two-photon resonance $\Delta_{2,\pm}\approx \pm 2.4$, which is crucial for applications of high-quality two-photon sources. As to the middle branch with $\Delta_c=0$, which exhibits a super-Poissonian photon statistics with $g^{(2)}(0) > 1$ and $g^{(3)}(0) > 1$. This result can be readily understood via the energy spectrum deduced from matrix (\ref{matrix}). For $\Delta_c=0$, the $n$-photon emission is resonant with an arbitrary photon excitation number $n$ since the energy splitting for the  dressed state $|n, 0\rangle$ is always equal to zero. Indeed, the super-Poissonian photon emission at the middle branch can be ascribe to the resonant multiphoton excitations of the cavity.

To demonstrate the two-photon bundles emission, we further calculate the typical $g_1^{(2)}(\tau)$ (solid line) and $g^{(2)}_{2}(\tau)$ (dashed line) versus the interval time $\tau$ at the red sideband of two-photon resonance $\Delta_{2,-}/g= 2.5$, as shown in Fig.~\ref{drivenatom}(f). Remarkably, the bundle-emission nature of the two-photon state is satisfied, where $g_1^{(2)}(0) > g^{(2)}_1(\tau)$ to ensure bunching for single photons and $g_2^{(2)}(0) < g_2^{(2)}(\tau)$ to ensure antibunching for separated two-photon bundles. Moreover, the lifetime of the bunching with respective to single photons and the antibunching for the separated two-photon bundles have the same time scale proportional to $1/\kappa$. We should note that our results reveal a new strategy to realize nonclassical photon emissions from high-quality two-photon bundles state to super-Poissonian dark state.

Compared to strong single-PB generated by driving the cavity field individually, the physical mechanism for realizing strong two-photon bundles emission can be ascribe to the interplay between $\delta_2$ enhanced energy spectrum anharmonicity and pump field $\Omega$ benefiting from the internal degrees of freedom in spin-1 atom. Indeed, the mechanism for generating two-photon bundles emission by using direct atomic transition $|g\rangle  \leftrightarrow |m\rangle$ induced by the classical pump field is analogue to apply a strong two-photon parametric driven, $\eta(\hat{a}^{\dag2}+\hat{a}^2)$. We should note that the tunable multiphoton parametric process requiring a high-order optical nonlinearitiy remains a challenge in experiments of cavity QEDs. More importantly, the high-quality two-photon bundles state is not relying on the strong atom-cavity coupling in contrast to the proposals in conventional spin-$1/2$ JCM with requiring $g/\kappa\sim 10^2$ ~\cite{munoz2014emitters,Deng:21}, which facilitates the experimental feasibility for engineering special nonclassical quantum states.

\subsection{Optical switching between single photon to two-photon bundles emission.}
\begin{figure}
\centering \resizebox{12cm}{!}{\includegraphics{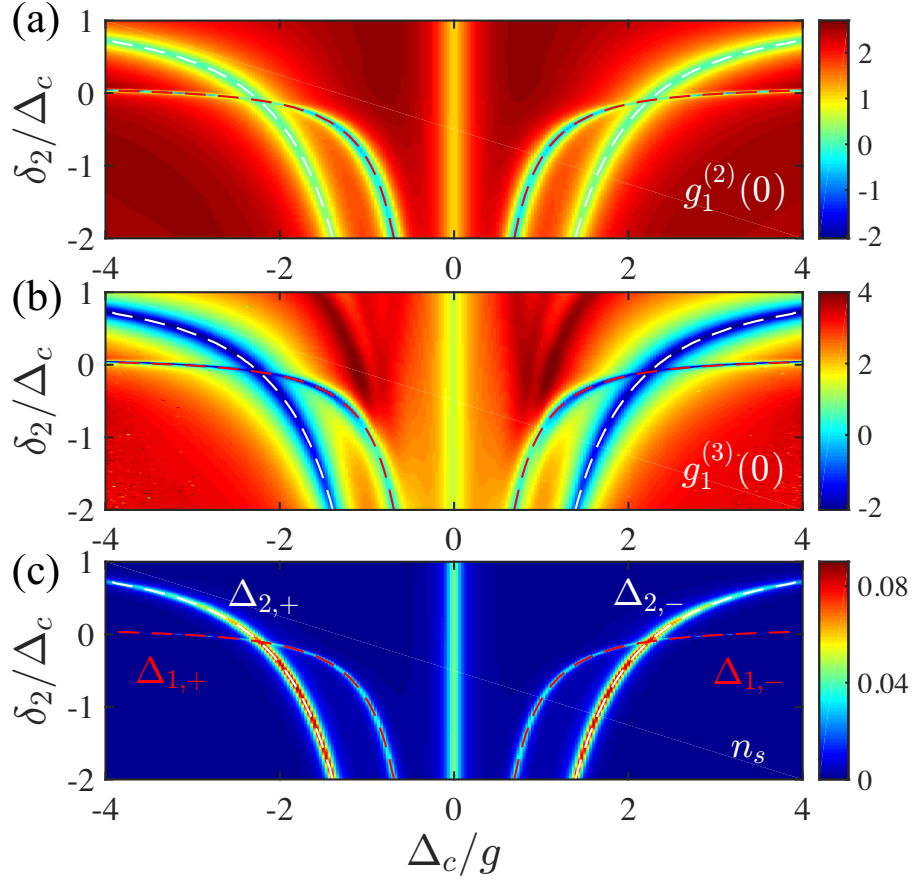}}
\caption{(color online). {\bf Optical switching from single photon to two-photon bundles.} Contour plots of (a) log[$g_1^{(2)}(0)$], (b) log[$g_1^{(3)}(0)$], and (c) $n_s$ as functions of $\Delta_c$ and $\delta_2$ for $\eta/\kappa=0.1$ and $\Omega/\kappa=0.08$. The white lines show the single-photon ($\Delta_{1,\pm}$) and two-photon ($\Delta_{2,\pm}$) resonances for the first and second dressed states of the energy spectrum.}
\label{drivenatomandcavity}
\end{figure}

We now turn to study the properties of the photon emission with combining the cavity driven and atom pump fields, simultaneously. Without loss of generality, we take the Rabi frequency $\Omega/\kappa=0.08$ and the cavity driven amplitude $\eta/\kappa=0.1$. We point out that the energy spectrum is roughly unchanged with these weak driven fields. Figure~\ref{drivenatomandcavity} summarizes the main results of the single photon to two-photon bundles emission in our system. In Figs.~\ref{drivenatomandcavity}(a) and \ref{drivenatomandcavity}(b), we map out the two-order and three-order correlation functions log[$g_1^{(2)}(0)$] and log[$g_1^{(3)}(0)$] as functions of $\Delta_c$ and $\delta_2$, respectively. In particular, the dashed (dashed-dotted)  lines depict the $\delta_2$ dependence of frequency for two-photon resonance $\Delta_{2,\pm}$ (single-photon resonance $\Delta_{1,\pm}$) calculated by diagonalizing $\hat{\cal H}_{I}$. Coexistence of single-photon and two-photon emissions is observed in a large parameter regime when $\delta_c/\Delta_c < 0.04$. Moreover, $\Delta_{1,\pm}$ and $\Delta_{2,\pm}$ merge to a single point when the level crossing between single-photon and two-photon resonances occurs at $\delta_2/\Delta_c =-0.1$.

With the further increasing of $\delta_2$, single-photon emission disappears since the value of $\Delta_{1,\pm}$ is divergence at $\delta_2/\Delta_c= 0.04$, which is highly consistent with the analytically result of energy spectrum in Fig.~\ref{scheme}(e). Interestingly, the purity of two-photon bundles emission can be further enhanced with increasing $\delta_2$, corresponding to $g_1^{(3)}(0)$ rapidly decreases with suppressing the multiphoton ($n>2$) excitations. In particular, as shown in Fig.~\ref{drivenatomandcavity}(c), the local minimum values of $g_1^{(2)}(0)$ and $g_1^{(3)}(0)$ in $\Delta_c$-$\delta_2$ parameter plane  corresponding to the large steady-state photon numbers $n_s$ for both single-photon and two-photon resonances. Moreover, the dark-state at $\Delta_c=0$ also exhibits a large photon emission and super-Poissonian photon statistics with $g_1^{(2)}(0)>1$.

In addition to requiring the antibunching for separated  photon bundles, the $n$-photon ($n>1$) bundles emission at $n$-photon resonance naturally possesses $n$-photon blockade satisfying $n$-photon bunching $g_1^{(n)}(0)>1$ and $(n+1)$-photon antibunching $g_1^{(n+1)}(0)<1$ for only characterizing the single photons. It is worthwhile to mention that there exists the quantum interference for photon statistics with interplay of the cavity driven and classical atom pump fields, which is similar to the construction of quantum interference conditions by employing both cavity driven and microwave fields for spin-$1/2$ JCM~\cite{Tang19}. Finally, our proposal of high-spin JCM can be used to realize nonclassical all-optical switching from the high-quality single photon to two-photon bundles and super-Poissonian photon emission by tuning $\Delta_c$ and $\delta_2$~\cite{Tang21, Volz2012, Chen768}, which may provide a broad physics community for applications in quantum information processing and quantum communication~\cite{munoz2014emitters}.

\section*{DISCUSSION}
Based upon the current experiments in atom-cavity QEDs, we theoretically investigate the strong single-PB to two-photon bundles emission in a trapped single spin-1 atom coupled to a single-mode optical cavity. We show that the energy-spectrum anharmonicity associated with $n$th dressed-state splittings can be significantly enhanced by tuning the
quadratic Zeeman shift in spin-1 JCM. The high-quality single-photon emission with $g_1^{(2)}(0)=7.5\times 10^{-5}$ and two-photon bundles emission with $g_2^{(2)}(0)=4.9\times 10^{-5}$ are achieved by driving the cavity and atom at a moderate atom-cavity coupling, respectively. Compared with the seminal spin-$1/2$ JCM, $g_1^{(2)}(0)$ in our scheme can be reduced by three orders of magnitude with a large cavity photon number. The strong optical nonlinearities are generated by the advantage of versatile spin degrees of freedom in high-spin systems, which helps to weaken the strong single atom-cavity coupling regime required by the seminal spin-$1/2$ JCM. In particular, the photon emissions can be highly tuned from the strong single-PB to two-photon bundles with interplay of cavity and atom driven fields, corresponding to the transition of strong antibunching for isolated photons to separated two-photon bundles. Furthermore, the scheme for realizing $n$-photon emissions with large steady-state photon numbers beyond the strong-coupling limit could be used as high-quality multiphoton sources. Our study offers an exciting opportunity for studying novel nonclassical quantum states by utilizing spin degrees of freedom in high-spin ($S\gg 1$) single atom-cavity QEDs, e.g.,  $^{173}$Yb with $S=5/2$~\cite{doi:10.1126/science.aaa8736}, which could provide a new building-block for exploring fundamental physics and applications in quantum computation~\cite{Pan2022, Rempe2022}. In the same spirit of our approach, we point out that our scheme can be further extended to explore new nonequilibrium phases and strongly correlated many-body physics in $N$-level Dicke model between atomic ensembles and optical cavity~\cite{PhysRevLett.127.033602,PhysRevLett.119.213601,zhiqiang2017nonequilibrium,PhysRevLett.122.103601,PhysRevA.104.063705,PhysRevLett.127.253601,PhysRevLett.128.153601}.

\section*{METHODS}
To investigate the photon quantum statistical properties for spin-1 JCM, we calculate the photon emission by numerically solving the quantum master equation using the quantum optics toolbox~\cite{tan1999computational}.  Taking into account the dissipations of cavity and atom fields, the Liouvillian superoperator $\cal{L}$ associating with the Lindblad-type master equation satisfy
\begin{equation}\label{master equation}%
{\cal{L}}\rho = -i [\hat{\cal H}_I+\hat{\cal H}_d, {\rho}] + \frac{\kappa}{2} \mathcal
{\cal{D}}[\hat{a}]\rho + \frac{\gamma}{2} \mathcal
{\cal{D}}[\hat{S}_{-}]\rho,
\end{equation}
where $\rho$ is density matrix of the single atom-cavity coupled system, $\mathcal {D}[\hat{o}]\rho=2\hat{o} {\rho} \hat{o}^\dag - \hat{o}^\dag \hat{o}{\rho} - {\rho}
\hat{o}^\dag \hat{o}$ denotes the Lindblad type of dissipation, and $\gamma$ is the atomic decay rate of the spin-1 ground state. Here $\hat{\cal H}_d=\Omega(\hat{\sigma}_{gm} +\hat{\sigma}_{mg})+\eta(\hat{a}^\dag+\hat{a})$ represents the weak separably controlled driven (pump) field for cavity (atom) field. We should emphasize that the weak pure dephasing effect for the ground states of atom field is safely neglected.

To identify the photon quantum statistics, the key physical quantity of time interval $\tau$ dependent generalized $k$th-order correlation function for photon emission is introduced as~\cite{munoz2014emitters,Deng:21}
\begin{align}
g_n^{(k)}(\tau_1,\ldots,\tau_n)=\frac{\left\langle \prod_{i=1}^k\left[\hat{a}^{\dagger }(\tau_i)\right]^n \prod_{i=1}^k\left[\hat{a}(\tau_i)\right]^n\right\rangle}{\prod_{i=1}^k\left\langle \left[\hat{a}^{\dagger }(\tau_i)\right]^n\left[\hat{a}(\tau_i)\right]^n \right\rangle},
\label{inequation}%
\end{align}
with $\tau_1\leq...\leq\tau_n$. We point out that the correlation function $g_n^{(k)}(\tau_1,\ldots,\tau_n)$ can be used to further characterize the $n$-photon bundles state in contrast to the standard $k$th-order correlation function $g_1^{(k)}(\tau_1,\ldots,\tau_n)$ which only for isolated photons as defined in Ref.~\cite{PhysRev.130.2529}. For zero time interval $\tau=0$, the equal-time $k$th-order correlation function $g^{(k)}_n(0)$ can be directly obtained by numerically solving the steady-state density matrix in Eq.~(\ref{master equation}) with ${\cal L}\rho_s = 0$, corresponding to the steady-state intracavity photon number ${n}_s = \langle \hat{a}^\dag \hat{a}\rangle$. Furthermore, the time-dependent multiphoton correlation function $g_n^{(k)}(\tau_1,\ldots,\tau_n)$ can be calculated by using quantum regression theorem with $\tau_1<\ldots <\tau_n$~\cite{carmichael2013statistical}.

The criteria of the correlation function for single-PB should satisfy two conditions: $g_1^{(2)}(0)<1$  to guarantee sub-Poissonian photon statistics and $g_1^{(2)}(0)<g_1^{(2)}(\tau)$ to guarantee photon antibunching. Similarly, the conditions for $n$-photon PB are $g_1^{(n)}(0) >1$ and  $g_1^{(n+1)}(0) < 1$  to ensure the $n$th-order super-Poissonian and $(n + 1)$th-order sub-Poissonian photon statistics simultaneously, and $g_1^{(2)}(0)>g_1^{(2)}(\tau)$ to ensure the photon-bunching for isolated photons~\cite{PhysRevLett.118.133604,PhysRevLett.121.153601}. For $n$-photon bundles state, an additional condition of $g_n^{(2)}(0)<g_n^{(2)}(\tau)$ is needed to ensure the photon antibunching between separated $n$-photon bundles~\cite{munoz2014emitters, Deng:21}. As to the experimental feasibility, the second-order correlation function $g_1^{(2)}(\tau)$ can be directly extracted via a Hanbury Brown and Twiss interferometer~\cite{Birnbaum2005}.  The experimental measurement of $g_2^{(2)}(\tau)$ can be achieved by using two-photon absorption in semiconductor photon detectors~\cite{Boitier:2009aa}.

Finally, the proposed $n$-photon emissions in spin-$1$ JCM can be realized in single ground $F=1$ alkaline-metal atoms, which has been demonstrated the capability in current experiments for single multi-level atom confined in a high-finesse cavity, e.g., $^{87}$Rb~\cite{mucke2010electromagnetically}. In our numerical simulation, the decay rate $\kappa=2\pi\times160$ kHz of the high-finesse optical cavity is set as typical energy unit of the system~\cite{sciadv.1601231}. We consider a single atom-cavity coupling $g/\kappa=6$ and a long-lived decay rate of the atomic ground state $\gamma/\kappa=0.01$. For convenience, the linear Zeeman splitting was fixed as $\delta_1/\Delta_c=0.1$. As a result, the tunable parameters in our system include light-cavity detuning $\Delta_c$, cavity driven strength $\eta$, Rabi frequency of the classical pump field $\Omega$, and quadratic Zeeman shift $\delta_2$. Without loss of generality, the strong PB of $k$-photon emissions is defined as $g_1^{(k)}(0)<0.01$.

\section*{DATA AVAILABILITY}
\noindent
The data that support the findings of this study are available from the corresponding author upon reasonable request.

\section*{ACKNOWLEDGEMENTS}
\noindent
This work was supported by the National Key R$\&$D Program of China (Grant No. 2018YFA0307500) and NSFC (Grant No. 12274473, Grant No. 12135018, Grant No. 11874433, Grant No. 11804409).

\section*{AUTHOR CONTRIBUTIONS}
\noindent
J.T. and Y.D. conceived the idea and wrote the manuscript. J.T. performed the theoretical and numerical calculations. J.T. and Y.D. supervised the project.

\section*{COMPETING INTERESTS}
\noindent
The authors declare no competing interests.

\section*{REFERENCES}
\noindent

\end{document}